\documentclass[sigconf]{acmart}

\usepackage{multirow}
\usepackage{longtable}

\AtBeginDocument{%
  \providecommand\BibTeX{{%
    \normalfont B\kern-0.5em{\scshape i\kern-0.25em b}\kern-0.8em\TeX}}}

\setcopyright{none}
\renewcommand\footnotetextcopyrightpermission[1]{}
\begin{document}

\title{A Structured Analysis of Information Security Incidents in the Maritime Sector}


\author{Monina Schwarz}
\affiliation{\institution{Universität Hamburg}\city{Hamburg}\country{Germany}}
\email{monina.schwarz@uni-hamburg.de}

\author{Matthias Marx}
\affiliation{\institution{Universität Hamburg}\city{Hamburg}\country{Germany}}
\email{matthias.marx@uni-hamburg.de}

\author{Hannes Federrath}
\affiliation{\institution{Universität Hamburg}\city{Hamburg}\country{Germany}}
\email{hannes.federrath@uni-hamburg.de}


\begin{abstract}
Cyber attacks in the maritime sector can have a major impact on world economy. However, the severity of this threat can be underestimated because many attacks remain unknown or unnoticed. We present an overview about the publicly known cyber incidents in the maritime sector from the past 20 years. In total, we found 90 publicly reported attacks and 15 proof of concepts. Furthermore, we interviewed five IT security experts from the maritime sector. The interviews put the results of our research in perspective and confirm that our view is comprehensive. However, the interviewees highlight that there is a high dark figure of unreported incidents and argue that threat information sharing may potentially be helpful for attack prevention. From these results, we extract threats for players in the maritime sector.
\end{abstract}

\maketitle
\pagestyle{plain}

\section{Motivation}
A large part of the global trade in goods is transported by sea. Thereby most cargo handling takes place in a few large ports world-wide. 
Composed of participants like logistic companies consisting of shipping companies and dockyards, ports are part of the maritime sector, which is part of the transportation sector. 
While providing opportunities for fast communication between distant participants, interconnectivity also offers new attack vectors for criminals to steal goods and data, sabotage systems and commit further crimes. 
For criminals, targets are particularly lucrative where they can get money or attract media attention, for example with critical infrastructure. 
Depending on size and sub-sector, companies are considered critical infrastructure. This includes, for example, ports and port logistics companies above a certain size~\cite{EU2008CriticalInfra,Homeland2015SectorPlan}. 
However, the outage of smaller companies should not be neglected, as the failure of many smaller participants can have the same severe consequences as well.  

Cyber attacks go through the media, such as the one on the Port of Antwerp~\cite{2013-BBC-Police_war} from a gang of smugglers or the big outtage of Maersk infrastructure in 2017~\cite{2018-Safety_4_S-Maersk_Lin}. 
The cyber incidents targeting the maritime sector reflect the diverse motivations of the attackers:
\begin{itemize}
\item weakening the economy of a country,
\item protesting against political decisions,
\item sabotaging opposition,
\item spying out information,
\item stealing money or cargo and
\item smuggling.
\end{itemize}
Cyber incidents will come to light under certain circumstances. German laws, for example, oblige critical infrastructure to report incidents to the authorities~\cite{BSIGesetz}.
According to EU's General Data Protection Regulation, organizations must communicate data breaches to supervisory authorities.
Ad hoc and other capital market notifications may also lead to publication of incidents.
Other reasons include that the impact is unconcealable or publication through third-parties including the attackers themselves, researchers or security companies. 
However, through personal experience and discussions with insiders, we assume that only a small subset of attacks on the maritime sector are publicly known. Companies could benefit from the incident knowledge of other players from the same sector. Nevertheless, most companies don't want to share or publish incidents, likely due to fear of bad publicity. 

To illustrate the threats to the maritime sector and give an incentive to share threat information, we show our threat taxonomy in section \ref{methodology}, present an overview about the publicly known cyber incidents from the past 20 years in section \ref{attacks} and evaluate our findings by conclude the results from expert interviews with insiders in section \ref{interview}. We complement this part with the related work in section \ref{relWork} and end with our conclusion in section \ref{conclusion}. 

\section{Related Work}
\label{relWork}
Security incidents in critical infrastructures are perceived more by the public than than attacks on non-critical infrastructure. This happens not least because of governmental reporting requirements for incidents in critical infrastructures. There are surveys and investigations targeting cyber incidents in the critical infrastructure~\cite{ogie2017cyber,al2018cyber, miller2012survey}. Others target cyber security in the maritime sector~\cite{mrakovic2019maritime}.   

\label{Shen}
The damage, that can occur in a scenario where more than one actor from the maritime sector is affected, was examined 2019: The so called Shen attack scenario~\cite{2019-CyRiM_Repo-Shen_attac} describes the impact of a computer virus infection of Asia-Pacific ports.  
In the scenario, a cargo management system is infected with a virus that spreads via traveling ships to other ports and their cargo management systems. As a result of the infection, these ports have to shut down their computer systems. Depending on the number of closed ports, 6 to 15, the estimated damage range between 80 billion USD and 219 billion USD.

Miller et. al~\cite{miller2012survey} classified SCADA incidents from the Repository of Industrial Security Incidents (RISI) according to the following classification factors: source sector, operation method, impact and target sector. They attempted to give a standardized and comparable terminology to determine incident severity levels.

Ogie~\cite{ogie2017cyber} also investigated SCADA and critical infrastructure related cyber incidents until 2014 using RISI data. He categorized the attacks according to the parameters intent, method of operation and perpetrator of the attack. By analysing 242 incidents, he gave a good overview of their distribution in different countries. 

Al-Mhiqani et. al~\cite{al2018cyber} targeted cyber incidents in cyber-physical systems in organization of Islamic cooperation countries until 2018. They showed the main intentions in the attacks according to the taxonomy: type of attack, target sector, intention, impact and incident categories. 

Mrakovic et. al~\cite{mrakovic2019maritime} gave an overview of the IT structure in vessels and international regulations and guidelines. Based on selected attacks, they showed digital risks that are specific to the maritime sector and presented maritime cyber risk management. 
There are case studies on maritime security in general. For example the Centre for Strategy \& Evaluation Services examined political context and security challenges of the maritime sector in a field study~\cite{csesCaseStudy11}. In 2011, D. Cimpean et. al analysed cyber security aspects in the maritime sector. They used different 
interviews and workshops to evaluate security and security measures in the maritime sector. They concluded, that the awareness of cyber security, the consideration of cyber security in regulations and a structured governance context are insufficient~\cite{cimpean2011analysis}. 

In~\cite{alcaide2020critical} participants of the maritime sector were interviewd by researchers to find out their involvement in cyber security and cyber incidents. Additionally, they included the maritime infrastructure, cyber threats, cyber attacks on board and international regulations into their analysis.\\
To the best of our knowledge, our study is the first that makes a summary of all currently published cyber incidents in the maritime sector and qualifies them through insider interviews.

\section{Threat Model and Methodology}
\label{methodology}

We have analysed publicly available data to find the most common types of attacks on the maritime sector in recent years. We considered common published successful attacks, prevented attacks and published Proof-of-Concepts (PoC) which are related to the maritime sector. Prevented attacks and PoCs are included in our statistic, because we wanted to create an overall picture of possible attacks. An unknown number of incidents stays unpublished. To estimate a real world distribution of attack vectors, we have to take potential attacks, i.e. PoCs and prevented attacks, into consideration. In order to organize the attacks, we divided them into categories described in section~\ref{threats}. 
These categories might still provide an incomplete view of the threat landscape, which is why we summarize our findings from this research in section~\ref{attacks} and compare them later to information from non-public sources \ref{interview}. In total, we gathered information on 90 different attacks and 15 PoCs.

\subsection{Digital Infrastructure}
The maritime sector can be divided into different types of infrastructure. 
Roughly summarized, there are ports that consist of the land part, which includes logistic companies and maritime service providers 
and the sea part which includes ship owners with supply industry and vessels. Looking at these participants' infrastructure, it consist of the following components:
\begin{itemize}
\item off-the-shelf components for everyday business (e.g. operating systems),
\item network infrastructure components, 
\item custom made operational technology (OT) like cargo bridges,
\item mixed information technology (IT) and OT systems like ships.
\end{itemize} 
The IT and OT systems in the land infrastructure are usually more or less intertwined. 
Ships, the most IT-isolated group among the sea component, used to have separate IT and OT on board and received new data only by rare manual updates. Nowadays, ships have on-board Internet and load new data like container information in every port via USB flash drives. The shipping crew may bring their own devices which connect with the ship's intranet which opens up to the internet. 
The IT and OT on ships and in the maritime sector in general are merging with each other, which means that more and more standard malware potentially can be found in every component.

\subsection{Threats}
\label{threats}
To bring the attacks analyzed in this paper into an order, we have used the 
threat taxonomy introduced by the European Union Agency for Cybersecurity (ENISA) in 2019
~\cite{enisa2019}. They derived sub-categories of threats out of port hazards. 
Even though attack goal, source of infection and other possible consequences may be linked back to different threats, we use main threats to group the attacks. The main threat categories are: 
\begin{itemize}
\item abuse and theft of data,
\item data manipulation,
\item denial of service,
\item geo-localisation signal spoofing and jamming,
\item interception of emission,
\item different malware-categories,
\item man-in-the-middle,
\item social engineering, phishing and identity theft, 
\item targeted attacks,
\item unknown attack vectors and
\item vulnerabilities exploitation of systems or devices.
\end{itemize}  

The threat categories may not always match completely with the threats caused by the attacks showed later on, as some of our sources may be inaccurate, incomplete or incorrect. In some cases, the threat category describes the attack itself and in others the effect because we often only know one of the two clearly from the reports. We never distinguish between the causes of infection because in most cases we do not know it.
For every attack, we create a profile with the following information about the details, as far as we know them: victim, perpetrator, year, threat category, threat name, state, damage sum, impact and source. All online sources are archived using the Internet Archive's Wayback Machine. 

\newcommand{\attackYear}[1]{\par\sffamily #1:\rmfamily~}  
\newcommand{\markPoC}{$^\dagger$}

\section{Threat Categories}
\label{attacks} 
In total, we gathered 90 attacks and 15 PoCs from different public sources. In the following we give a short description and one source per event, grouped by threat category and ordered by year. We categorized the attacks according to the main threat, although this is sometimes not clearly possible due to the fact that we do not know the way of intrusion, cause and impact about them. We marked all PoCs with the follwing symbol: $\dagger$.

\subsection{Abuse and Theft of Data}
\label{abuseTheft}
For various reasons data from maritime participants can be valuable for attackers. When hackers steal sensitive data or abuse certificates used in port operations like customs, the availability and integrity of these operations can be violated. In most cases, it was not clear how the attackers got the data, whether by malware, credential theft or by other attack vectors.
\attackYear{2012}Attackers gained access to systems operated by the Australian Customs and Border protection which allowed the attackers to check whether their shipping containers were regarded suspicious by police or customs~\cite{2018-Yachting_P-Could_you}. 
\attackYear{2013}Designs of the Littoral Combat Ship from the US Navy and other weapon systems have been stolen. Chinese hackers were suspected of having committed the crime~\cite{2013-The_Washin-Confidenti}.
\attackYear{2014}A person onboard a vessel had been uploading images on Facebook, which provided a detailed look into the safety measures in place. This information leak could have led to attackers gaining information about the ship, which would make it easier to attack. This was discovered prior to entering the Gulf of Aden and the vessel changed its course because they feared an attack from pirates~\cite{2014-Cyberkeel-MARITIME_C}. 
\attackYear{2016} The ISC-CERT warned about a SQL injection vulnerability under active exploitation affecting web applications used by multiple US organizations that could lead to data loss~\cite{2016-ICS-CERT-Navis_WebA}. 
South Korea suspects North Korea of having hacked Daewoo Shipbuilding Maritime Engineering (DSME) and stealing blueprints of warships~\cite{2017-Reuters-North_Kore}. DSME may be victim of another data theft in 2020, but the hack is not further described and still unconfirmed~\cite{2020-Twitter-Threat_act}. 
Also, the personal data of thousands of US Navy soldiers were stolen by infiltrating a service employee's notebook~\cite{2016-Reuters-Personal_d}. 
Information on the details of six submarines designed for the Indian Navy was stolen from the Naval Group, a french defence contractor formerly known as DCNS. In total they lost over 22,000 pages containing confidential material~\cite{2016-Reuters-Frances_D}. 
Pirates infected the content management system of an unnamed company to gather information about which container on a vessel is valuable and could be robbed~\cite{2016-businessin-High-tech}. 
\attackYear{2017}Svitzer Australia detected that they leaked data through the email accounts of three employees. The stolen data contained personal information about the company's employees~\cite{2018-ABC_News-Svitzer_em}. 
Clarkson was the target of a cyber security incident, where an unknown attacker accessed internal systems and stole data. The attacker demanded ransom for the return of the stolen data~\cite{2018-Clarksons-Update_on}. 
An Australian Defence subcontractor lost several gigabyte of sensitive documents, related to allied defence projects like jet fighters and ship plans~\cite{2017-iTnews-Hacked_Aus}. 
\attackYear{2018}Early in the year, a US Navy subcontractor was attacked and 64 gigabyrte of data including confidential plans for supersonic missile projects were stolen. The USA suspected the Chinese government of the hack~\cite{2018-The_Telegr-Chinese_ha}. 
Attackers attempted to sell material of defence ship builder Austal on the Internet. Austal said they do not intend to respond to that kind of ransom and thereby risked the leak of the stolen data~\cite{2018-The_Guardi-Defence_sh}. 
\attackYear{2019}Holland America Line noticed unauthorized third-party access to email accounts containing personal information including social security numbers, health-related information and financial account information. The attackers likely got acces via phishing emails~\cite{2020-prnewswire-Holland_Am}. 
A group named ``Maze'' hacked London Offshore Consultants and asked them for ransom after a small selection of the stolen files were leaked online. The hackers claimed that they stole 300 GB of information but LOC confirmed, that only a small number of clients were affected~\cite{2020-Splash-London_Off}.
\attackYear{2020}Unconfirmed sources described that South Korea's shipbuilder DSME was hacked by unknown hackers that sold access to DSME's network~\cite{2020-Twitter-Threat_act}. 
\attackYear{2021}Attackers gained access to some of the internal systems from Carnival Corporation and accessed data relating to employees and guests of the company~\cite{2021-Bleeping_C-Carnival_C}. 

\subsection{Data Manipulation}
Shipping companies rely on their data sources about vessels and customer information. Ships need reliable nautical data to navigate. Manipulating data unnoticed or even visibly is attractive for attackers to achieve goals including: finding valuable goods, sabotage companies, or deface companies.

\attackYear{2014}The website of the Massachusetts Maritime Academy was presumably hacked by an Islamist group. They replaced the content of the website with Islamic messages. It is presumed that the Moroccan Electronic Islamic Union group is responsible for hacking the website~\cite{2014-CyberKeel-Marine_Cyb}.

\subsection{Denial of Service} 4
\label{DDoS}
The aim of Denial of Service (DoS) attacks is to affect the availability of a system. Several sources can be involved in the same DoS attack which then can be called a distributed DoS (DDoS) attack. DDoS attacks are used for blackmailing but also as a political weapon~\cite{mansfield2015growth}. 

\attackYear{2001}The Port of Huston was attacked twice~\cite{2003-The_Regist-UK_teenage}: The first attack in August was a DDoS attack from a US citizen. The second attack was distributed ping flood attack, conducted in September by a US teenager who used the Port of Houston as an instrument in a personal revenge plan. 
\attackYear{2013}The Port of Long Beach was a target of large scaled DDoS attacks~\cite{2018-cytegic-Managing_t}. 
\attackYear{2017}The Port of Vancouver became a target. When asked for details, Vancouver port's spokeswoman said: \textit{``We cannot answer this, as it could pose a security risk for the port''}~\cite{2017-Columbian-Port_of_Va}.

\subsection{Geo-localisation Signals Spoofing and Jamming}
\label{spoofing}
Geo-localisation signals (GPS) and navigation systems can be spoofed or jammed in order to hinder navigation and change the trajectory of a vessel~\cite{enisa2019}. In the following, GPS spoofing attacks related to the maritime sector are listed. 
\attackYear{2001}A vulnerability assesment of the U.S. transportation infrastructure discusses attacks on GPS related to the maritime sector~\cite{carroll2003vulnerability}.\markPoC 
\attackYear{2009}To identifying the full effects of GPS jamming on safe navigation at sea, researchers from the General Lighthouse Authorities of the United Kingdom and Ireland examined the effects from unintentional interferences and jamming onto GPS~\cite{2009-Journal-GPS_jammin}.\markPoC 
\attackYear{2010}South Korea reported GPS jamming by North Korea~\cite{GPSVuln13}.
\attackYear{2012}South Korean officals said that 254 ships experienced GPS interruptions over 16 days of jamming by North Korea~\cite{2013-Reuters-Government}.
\attackYear{2013}Researchers from the University of Texas at Austin successfully demonstrated GPS spoofing for a super yacht at sea. With small custom-made GPS devices, they tricked the yacht onto a track hundreds of meters from its intended one. They also showed, that this attack could be performed silently~\cite{2013-Cockrell_s-UT_Austin}.\markPoC  
Trend Micro researchers showed multiple ways to disturb and spoof ships' AIS systems. They achieved modification of ship details, creating nonexistent vessels, trigger collision warnings, create nonexistent search and rescue helicopters, spoof the weather data and impersonate maritime authorities to trick the vessel~\cite{2014-Cyberkeel-MARITIME_C}.\markPoC 
Targeting more than GPS, a research team from IOActive described attacks on satellite communication hardware on ships. They examined devices from different vendors and found vulnerabilities in each of them. In result, they could cause a manipulation of communication structure between ships~\cite{2014-Technical_-A_wake-up}.\markPoC
\attackYear{2016}More than 700 ships were affected of GPS jamming over a period of five weeks. The signals originated from five North Korean locations along the border to South Korea~\cite{2016-Reuters-South_Kore}.
\attackYear{2017}All at once, the GPS navigation of more than 20 vessels was spoofed at the same time in the Black Sea near Russia. The effect lasted for a few days with different phenomena like loss of position and wrong display of positions.~\cite{2017-The_Mariti-Mass_GPS_S}.
\attackYear{2018}Another far reaching GPS spoofing was recorded between 2018 and 2019 in the Central and the Eastern Mediterranean Sea. The loss of GPS signal obstructed the functions of (GPS-based) navigation devices~\cite{2019-US_Depar-MSCI_Advis}. 
A proof of concept attack showed how specially crafted malware can be used through various channels such as USB-drives to alter data and attack the ECDIS navigation system in Ships~\cite{2018-Riviera-Bridge_sys}.\markPoC  
Researchers from the German police targeted both, the ECDIS and GPS Systems of Ships, to show how vulnerable the navigation is. The GPS is spoofed with cheap technical material and the ECDIS maps are faked and shared to alter the position of objects and the terrain information~\cite{2018-Wilfried_H-Polizei-In}.\markPoC  
\attackYear{2019}At the Port of Shanghai, ships Automatic identification systems (AIS) and GPS systems were spoofed. The AIS showed other ships on wrong positions and they appeared to congregate in large circles. The origin of these dissruptions is still unclear~\cite{2019-MIT_Techno-Ghost_ship}.

\subsection{Interception of Emission}
\label{interception}
For many reasons, hackers would like to intercept the communication between stakeholders like port and vessels. Often such interceptions are used to trick victims into sending money to the wrong target.
\attackYear{2014}The World Fuel Services (WFS) fell victim to a scam attack where they lost 18 million USD. The attackers tricked WFS to buy and deliver 17\,000 metric tons of gas to alledged buyers~\cite{2014-Ship_and_B-WFS_In_Cou}.

\subsection{Malware} 
\label{malware}
Since electronic components with common operating systems are part of ships and harbour control infrastructure, they can be infected with malware that also infects company systems in other sectors. The abilities and damage skills of malware vary and are therefore divided into different sub-categories. Ransomware will have its own part as its relevance has increased significantly in recent years. 

Also a lot of malware infections are part of the attack vectors from targeted attacks in section \ref{targeted}
therefore we do not mention them again here. Most of the malware incidents we found were caused by ransomware. In some other cases, too little information is known to be able to say whether ransomware or other malware was the cause.

Ransomware encrypts the targeted system and promises a decryption in exchange for a ransom fee. Some offshoots additionally extract data or only encrypt, without the opportunity to decrypt the data.

\attackYear{2013} A smartphone was connected to the network of a drilling unit, resulting in a not further described malware infection and a shut down of the machine~\cite{2015-Maritime_C-Coast_Guar}.
\attackYear{2017}A famous victim of the ransomware-like, destructive malware NotPetya is the Maersk Group. The destructive infection was probably not targeted but cost the company over 300 million USD~\cite{2018-Safety_4_S-Maersk_Lin,wired2018NotPetya}. 
Other NotPetya victims are the Port of Rotterdam~\cite{2017-NL_Times-Rotterdam}, Jawaharlal Nehru Port, Indias largest container port~\cite{2017-Healthcare-Nuance_kno}. 
\attackYear{2018}Cosco Shipping fell victim to NotPetya as well~\cite{2018-Safety_4_S-Cyber_atta}. 
The Port of San Diego was hit with the SamSam ransomware and lost more than 30\,000 USD as they refused to pay the ransom~\cite{2019-San_Diego_-What_happe}. An email campaign spreading multiplatform Java Remote Access Tool (RAT) took place~\cite{2019-Yoroi-The_Story}. This attack is simmilar to the MartyMcFly campaign, see section~\ref{Marty}. 
\attackYear{2019}Early that year, a vessel near the Port of New York was infected with Emotet that rendered much of its network unusable~\cite{2019-James_Rund-Coast_Guar}. A not further named MTSA (Maritim Transportation Security Act) regulated facility was infected with Ryuk ransomware. The ransomware most likely entered the company's network via phishing email~\cite{2019-US_Coast-Marine_Saf}. Ryuk also infected Pitney Bowes~\cite{2019-Security_W-Pitney_Bow,2019-Pitney_Bow-System_Upd}. A research team created and studied the Shen attack scenario~\cite{2019-CyRiM_Repo-Shen_attac} that we described in section~\ref{Shen}.\markPoC 
\attackYear{2020}The ship management company Anglo-Eastern partly prevented a ransomware attack. According to an Anglo-Eastern manager, only 20 percent of the workstations and 10 percent of the servers had been encrypted~\cite{2020-Splash-Anglo-East}. Toll Group was targeted by two attacks: The first was a large scaled attack with Mailto/NetWalker and the second occured with Nefilim. While there is no statement, how NetWalker infected the companies IT, Nefilim malware was most likely distributed via Remote Desktop Service~\cite{2020-CISOMAG-Mailto_Ran,2020-IT_News-Toll_Group,2020-IT_News-Toll_Group2}.
The Blue Water Shipping Company achieved to stop a ransomware attack taking place in September before the ransomware could encrypt major parts of the IT infrastructure~\cite{2020-ShippingWa-Blue_Water}. 
MSC Cargo was hit by an unnamed malware that caused the company to shut down servers and stop part of their operation for a few days~\cite{2020-Security_W-Shipping_G}. 
The cruiser line operator Carnival was infected with a not further specified ransomware. The infection could have happened through vulnerable end devices and accessed certain files~\cite{2020-Carnival-Substitute,2020-Bleeping_C-Worlds_la}. The Ragnar Locker ransomware hit CMA CGM. Distributed via email attachments the malware encrypted and exfiltrated data~\cite{2020-SWZ_Mariti-CMA_CGM_su}.
The Norwegian shipping company Hurtiguren has been affected by a ransomware that has infected and paralysed some of its systems that are part of its global infrastructure~\cite{2020-Security_W-Norwegian}.
\attackYear{2021}Through a prepared fake email, attackers delivered the so called PortDoor backdoor to the submarine defence contractor Rubin Design Bureau. According to researchers~\cite{2021-The_Hacker-New_Chines}, the attack is Chinese state-sponsored.

\subsection{Man-in-the-Middle}
In a Man-in-the-Middle (MitM) attack, an attacker inserts himself into a communication and pretends to one or both communication partners that they are talking directly to the other. Thereby the attacker can normally read and modify the communication.
\attackYear{2013}Clearsky analysts in cooperation with CyberKeel examined a bunkering scam attack from 2013 and 2014. They found that more participants of the maritime industry are vulnerable to MitM attacks in their business communication and could therefore fall for bunkering scam attacks~\cite{2014-Cyberkeel-MARITIME_C}.\markPoC 
\attackYear{2015}CyberKeel researchers found that websites of carriers are vulnerable to a faked-website spoofing. With a perfect copy an attacker could trick shipper into entering their sensitive information into the faked website and then pretend to be the shipper~\cite{2015-Lloyds_Loa-Top_contai}.\markPoC

\subsection{Social Engineering, Phishing and Identity Theft}
\label{phishing}
Through social engineering, attackers may trick people into giving away confidential information. This includes phishing where attackers use email or websites to pretend to be someone else. So called spear phishing attacks are tailored to the victims by imitating familiar or trustworthy addresses and names. 
\attackYear{2013}Messages with remittances where sent from a Singapore shipping company to its bank from the companies managing director's email account but without his knowledge. There were four transactions of in total 1.8 million USD~\cite{2018-Kennedys-Whaling}. 
\attackYear{2014}The maritime cyber security company CyberKeel tested the eCommerce applications of different shipping carriers and found weaknesses in the password policies what may allow them to bruteforce and access customers accounts~\cite{2014-Cyberkeel-Maritime_C2}.\markPoC 
\attackYear{2015}Hackers tricked a shipping company based in Cyprus to pay the money for a legitimate ongoing supply to a wrong banking account and stole around 645\,000 USD~\cite{2016-American_a-MARITIME_C}. 
The Nautilus Minerals company lost 10 million USD to an unknown party while thinking they paid a charterer's guarantee to the Marine Assets Corporation. The exact reason for this failure is unpublished~\cite{2015-Insurance_-Cyber_hack}. 
\attackYear{2016}A seaman fell victim to a honey trapping attack: A blackmailer lured him into taking suggestive pictures of himself and sent them to the attacker over a period of six weeks. Later the attacker demanded a deposit of 10\,000 USD to not show the pictures to his family. Researchers warned that it seems like this kind of attack happens more often in the Gulf of Guinea region~\cite{2016-Paul_Berri-Honeytraps}. 
A broad email fraud campaign spotted by Clearsky Security targeted VersaCold and Toll Group amongst others with real-looking and targeted fake domains~\cite{2016-ClearSky-Business_E,2016-American_a-MARITIME_C}. 
\attackYear{2017}A Malaysian bunker fuel owner became victim of a fraud where he lost 4.5 million Malaysian Ringgit to an unknown attacker. To make the scam as credible as possible, the attacker obtained copies of the victim's email communication via spyware to masquerade a fake email~\cite{2017-The_Sun_Da-Kedah_bunk}. 
Between 2017 an 2018, the business email spoofing group Gold Galleon, likely from Nigeria, targeted maritime actors and stole a minimum of 3.9 million USD. They used among others tools for keylogging and password-stealing to gain access to email accounts from where they carried out their attacks~\cite{2018-Securework-Gold_Galle}. 
\attackYear{2019}The US Coast Guard warned about phishing emails and malware intrusion attacks targeting commercial vessels. The attackers tried to gain sensitive information by disguising themselves as official Port State Control~\cite{2019-US_Coast-Marine_Saf2}. 
\attackYear{2020}The Maritime Transportation System (MTS) and the Information Sharing and Analysis Center (ISAC) warned Tug owners about phising emails containing fake voicemail attachments~\cite{2020-Cybersecur-Cyber_Thre}.
The US Coast Guard warned in September about sophisticated email spoofing attacks which lead to at least one network compromise. The campaign used among others a faked Cost Guard address~\cite{2020-US_Coast-Marine_Saf}.
\attackYear{2021} During payment of a contract from the subsidiary Hammonia Tanker Holding to the parent company Hammonia Schiffsholding, the email traffic between the companies was manipulated in such way that the payments in the amount 1 million USD went to the attackers~\cite{2021-HAMMONIA_S-Finanziell}. 
The International Transport Intermediaries Club (ITIC) warned of an increased incidence of email fraud, citing a recent incident as an example. In this case a shipbroker contacted his usual bunker supplier but emails with faked payment details were sent by attackers, whereupon the 300.000 USD were transferred to the wrong account~\cite{2021-ITIC_Press-Counterfei}. ITIC also presented another case where a ship manager had arranged repairs with a shipyard. After a fake email, the payment of 500.000 USD was sent to the attackers instead of the shipyard~\cite{2021-ITIC_Press-The_phone}. The Red Sky Alliance warned of email frauds in which the attackers are impersonating the Mediterranean Shipping Company (MSC) to spread faked emails with dridex malware~\cite{2021-SAFETY4SEA-Recent_mar}.  

\subsection{Targeted Attacks}
\label{targeted}
Targeted attacks use infection strategies and attack vectors tailored to one specific target and are usually scheduled over a longer period of time. Often a combination of social engineering, phishing and public knowledge about the target is used for infection with a broad amount of tools ranging from data theft tools and remote-access Trojans to destructive malware.
\attackYear{2011}The Iranian Shipping Line IRISL fell victim to a huge hacking attack where its cargo data systems were damaged. Israel was accoused of having committed this attack. The motivation of the attacker was to discourage all people transporting material for the Iran atom program~\cite{2012-Reuters-Irans_top}.
Icefog is an Advanced Persistent Threat (APT) targeting maritime and shipbuilding groups and other industries in Japan and South Korea since at least 2011. The attackers relied on spear-phishing emails that attempt to trick the victim into opening malicious attachments or websites~\cite{2013-Kaspersky-The_ICEFOG}. From 2011 to 2013, the Port of Antwerp was target of a persistent targeted attack from drug smugglers, where they gained remote access to a terminal system that gave them the opportunity to access, steal and erase containers out of the shipping companies system~\cite{2013-BBC-Police_war}. 
\attackYear{2012}Danish maritime shipping authorites fell victim to a possible Chinese state sponsored attack with unnamed methods, possibly a virus infected pdf file. The goal seemed to be espionage~\cite{2014-ShippingWa-Kina_hacke}. The operation Cleaver, a probably Iranian worldwide cyberwarfare operation between 2012 and 2014 targeted different unnamed transportation companies. The infection contained compromising of active directory domains, Cisco Edge switches and routers and networking infrastructure~\cite{2014-Cylance-Operation}. 
\attackYear{2013}The so called Kimsuky APT group targeted mostly South Korean organisations, among them the Hyundai Merchant Marine, with malware. A dynamic link library was used as dropper for further infection and espionage on the infected system~\cite{2013-Kaspersky_-The_Kimsu}. 
\attackYear{2014}The so called Zombie Zero Attack was a broad ranged targeted attack onto shipping companies. It was a supply chain attack, barcode readers were delivered with embedded malware that was able to open a backdoor for further attacks in which finance data was stolen~\cite{2014-TrapX_Secu-Anatomy_of}. 
The hacker group named Leviathan, also known as TEMP.Pericsope, launched first attacks onto the maritime sector. One campaign was seen between November and January of the following year. The group faked emails with malicious Microsoft Office documents to US universities with military interests~\cite{2017-Proofpoint-Leviathan}. 
\attackYear{2015} More malware campaigns by the threat actor Leviathan were reported. Some victims were connected to South Chinese Sea issues~\cite{2015-Spiceworks-Stealthy_C, 2017-Proofpoint-Leviathan}.
\attackYear{2017}Early 2017, a German container ship between Cyprus to Djibouti was hacked, most likely by pirates. They took full control over the navigation system to enter the ship~\cite{2017-Safety_at_-Shipping_m}. 
\label{Marty}%
\attackYear{2018}Different enterprises, among them Italian naval industries, where targeted by an unknown group that deployed the remote-access Trojan MartyMcFly and the remote access tool QuasaRAT through specially crafted and targeted phishing emails~\cite{2018-Yoroi-Cyber-Espi}.
End of 2018, Russia seized Ukrainian ships. Before the attack, there were multiple coordinated cyber attacks on Ukrainian government agencies tied to Russia. One of these attacks was a phishing campaign, with the aim of installing malware, exfiltrating data and taking control over computer functions. In another attack in November 2018, Russian-linked Gamaredon Group delivered the windows tailored backdoor Pterodo to Ukrainian government agencies. Presumably the malware should have gathered information for the further attack where Russia seizured the Ukrainian ships~\cite{2018-Nextgov-Russia_Lau}. 
Once more, attacks by Leviathan were noticed. Amongst others, victims were an US shipbuilding company and, again, maritime entities connected to South China Sea issues. The threat actor spread their malware through spearphishing emails with Microsoft Office documents and installed multiple backdoors and spyware tools to gather data from the infected systems~\cite{2018-FireEye-Suspected}. 
\attackYear{2019}In Kuwait, at least one transportation and shipping organization was victim of the xHunt Campaign. The attackers installed a backdoor called Hisoka that gave them the opportunity to load other exploitation tools onto the infected system~\cite{2019-Paloalto_N-xHunt_Camp}. 
\attackYear{2020}Iran claimed that a large scale targeted attack was carried out on the infrastructure of the country's ports. No further details have been published so far~\cite{2020-Reuters-Iran_says}. 
Similar to the famous case in Antwerp, in Hamburg a South American drug cartel smuggled containers of cocaine. For this, an attacker with insider knowledge hacked into the IT systems of a container company and caused the containers to be delivered unnoticed~\cite{2020-Spiegel_on-Polizei_un}. 

\subsection{Unknown Threats}

In some cases, only the damage of the attack or the fact that an attack took place are known. This can be triggered by the fact that the attackers removed all evidence, the attack was not an attack but an incident, the system monitoring is too incomplete to reconstruct an attack or the company hides the investigation results for some reasons. In most cases, the minimal damage is a temporary shut-down of servers and other IT infrastructure. 
\attackYear{2015}During the process of finding a new submarine shipbuilder for Australia, the three competitors Germany’s TKMS and builders from Japan and France received massive hacking attempts. That forced them to carry out the deliveries of sensitive information by hand~\cite{2015-Business_I-Foreign_bu}.
\attackYear{2017}After an attack with unpublished attack vector, in which the attacker had access to intranet systems, the BW Group had to shut down multiple servers~\cite{2017-SP_Global-Shipping}. 
\attackYear{2018}An attack hit the Port of Barcelona but was repelled by internal security meassurements~\cite{2018-Ara-El_port_de}. 
\attackYear{2019}An unknown intrusion forced the UK-based marine services provider James Fisher and Sons to shut down affected systems after recognizing unauthorized access~\cite{2019-The_Mariti-James_Fish}. 
\attackYear{2020}The Shahid Rajaee Port of Bandar Abbas in Iran was the target of an unspecified attack. Local authorities told that they could repell most of the attack~\cite{2020-Washington-Officials}. 
The U.N. International Maritime Organization urged that they became the victim of an not further specified attack~\cite{2020-Reuters-UN_shippin}.
\attackYear{2021}The offshore vessel owner Bourbon was victim of an unspecified attack that affected and disabled parts of its internal computer network~\cite{2021-Splash247-Bourbon_co}. Hyundai Merchant Marine (HMM) confirmed an unidentified security breach, impacting some of the company's email server. HMM stated that no data had been leaked through the attack~\cite{2021-Splash247-HMM_confir}. 

\subsection{Vulnerabilities Exploitation of Systems or Devices}

Unpatched security vulnerabilities, weak passwords and 
wrongly configured systems offer a broad attack surface. A lot of systems have known vulnerabilities, although patches for these have often been available for a long time. If an attacker knows about these vulnerabilities, all he has to do is find systems that are affected by them. Shodan is one of the tools to simplify the search for vulnerable systems worldwide.
\attackYear{2013}A US Navy team found vulnerabilities in the USS Freedom while pentesting the vessels network~\cite{2013-Reuters-Cyber_vuln}.\markPoC 
\attackYear{2014}The US Naval Sea Systems Command warned that parts of vessel control systems, like a submarine's Caterpillar-build backup diesel  engine, running outdated and vulnerable Windows XP versions with known vulnerabilities that could be abused to break into the system~\cite{2014-Cyberkeel-Maritime_C2}.\markPoC 
\attackYear{2015}Microsoft patched a vulnerability that could allow attackers to take control of webservers. CyberKeel tested 50 different maritime websites against this vulnerability. 37 percent of the tested websites were vulnerable, beneath them websites from MSC, Happag Lloyd and Hamburg Süd~\cite{2015-Splash-Lack_of_pa}.\markPoC 
\attackYear{2017}A researcher from Pen Test Partners showed the take over of control over a ship's network by means of a vessel heading from Malacca Strait to Tubarao in Brazil. With the help of Shodan he found access to a VSAT modem that gave him information about the crew on board and the network configuration. This got him further into the ship's network~\cite{2017-Pen_Test_P-OSINT_from}. Another security researcher named x0rz told that they had hacked into a ship and gained control of the ship's VSAT systems. He said that they could see the call history, change system setting and upload new firmware. In addition, as in the previous attack, the access to the VSAT system leaks information about the ship's position~\cite{2017-Mfame-Remote_Att}.
\attackYear{2020}Researchers from Yangosat gained access to a vessel's VSAT system near a European capital city. They had access to information about the system and would have been able to break the vessel's  system completely~\cite{2020-Ewan_Robin-Hacked_-_a}.

\subsection{Conclusion of Attacks}
\label{targetConclusion}
\begin{figure}[ht]
\centering
\includegraphics[width=\columnwidth]{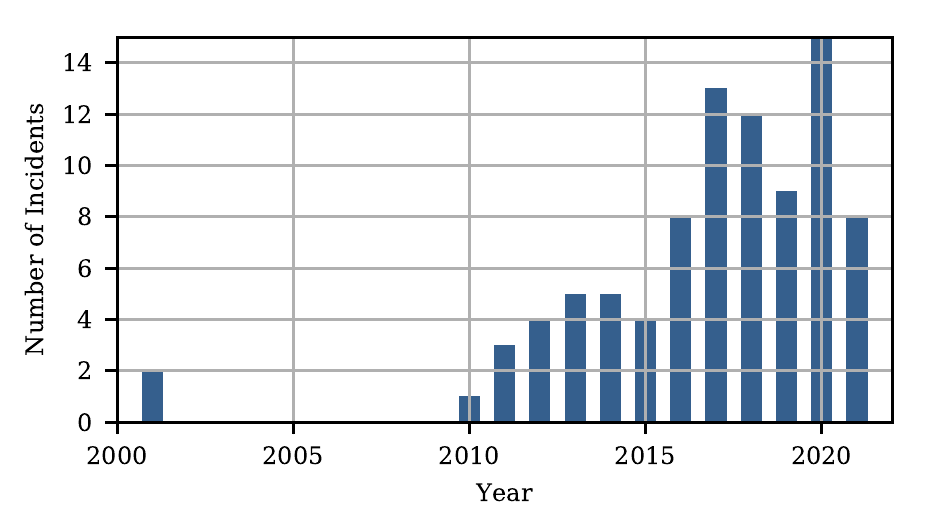}%
\caption{Number of publicly reported incidents per year.}
\label{fig:years}
\Description{The figure shows the number of publicly reported incidents per year. 2 inicidents were reported in 2001. Since 2010, the number of incidents is increasing.}
\end{figure}

\begin{figure}[ht]
\centering
\includegraphics[width=\columnwidth]{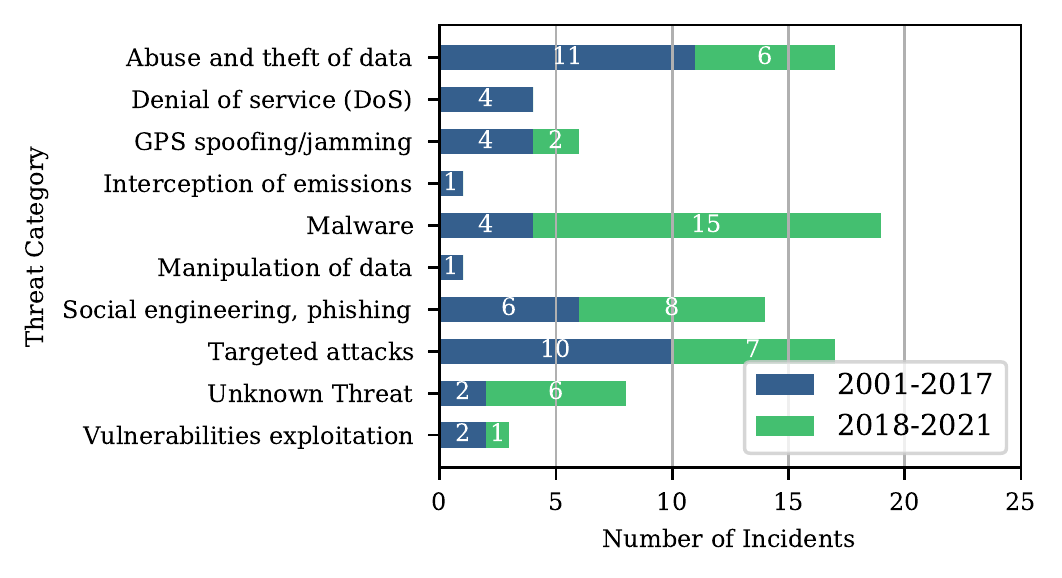}%
\caption{Number of publicly reported incidents per threat category for the periods 2001-2017 and 2018-2021.}
\label{fig:threats}
\Description{The figure shows the number of incidents in different threat categories for two periods. In recent years, most incidents were malware incidents.}
\end{figure}

The most common detected attacks onto the maritime sector in the last years are ransomware-based infections. The software used is similar to the software used in attacks on other sectors and also the path of infection is most common, via email. A lot of other malware infections, affecting shipping companies and systems on vessels, are standard malware infections with non-specialized software. There are also examples of detected per-attack-cropped software in the last years. One example for these targeted campaigns is email fraud, with shipping-company specific fraud email addresses and content. 
In addition, participants of the maritime sector were repeatedly affected as collateral damage by broad targeted malware attacks on system-relevant infrastructure.
By comparison with figure~\ref{fig:threats} which shows the number of incidents for the periods 2001 to 2016 and 2017 to 2021 per threat category, and figure~\ref{fig:years} which shows the distribution of attacks over the years, we see a clear trend towards an increase in numbers of incidents, especially in ransomware. A list of all incidents can be seen in the appendix table~\ref{tab:incidents}.

If one looks at the accused attackers, we find that global conflicts between states~\cite{globalconflicttracker2020} are reflected there: North Korea was accused of being behind cyberattacks on South Korean companies~\cite{2013-Reuters-Government,2017-Reuters-North_Kore}, Russia was accused of attacking Ukrainian targets~\cite{2018-Nextgov-Russia_Lau} and China allegedly attacked various targets in the conflict over the South China Sea~\cite{2018-Bloomberg-Chinese_Ha}.

\subsection{Recommended Sources for Further Information}
During our research, we looked through more than a hundred references. In this section, we want to highlight references that are particularly detailed, entertaining or educational and well worth reading.
Some of them are summaries from IT security companies that regularly present their research and working experience. Others are reports from security researchers or (former) employees. 

We would like to recommend two sources about the NotPetya incident at Maersk. First, the Wired article describing the incident in detail~\cite{wired2018NotPetya} is worth mentioning. Second, a report by Gavin Ashton, former Maersk employee, is remarkable. His report also covers a period prior to the incident~\cite{2020-Gavin_Asht-Maersk_me}. 
The Center for Advanced Defense Studies published a detailed and nicely prepared report on exposing GPS spoofing in Russia and Syria~\cite{2020-C4ADS-Above_Us_O}. 
There are a lot of extensive reports about APTs and specific campaigns. The APTnotes repository~\cite{APTnotes} collected more than 500 publicly available reports related to APTs. The repository is not specific to the maritime sector. As an example, we mention the reports on the likely Nigerian threat group Gold Galleon by Secureworks~\cite{2018-Securework-Gold_Galle} and on the Kimsuky operation by Kaspersky~\cite{2013-Kaspersky_-The_Kimsu} because of their length and level of detail.

\section{Expert Interviews}
\label{interview}
The set of cyber incidents that we collected shows only a fraction of all incidents that occured, as some incidents have not been detected at all~\cite{fireeye2020report} and neither the affected company, nor the attacker or a third party has disclosed the incident to the public
~\cite{rsm2019report,bsi2019report}. Furthermore, we might have missed publicly reported attacks.

To set our results in relation, we conducted five interviews with experts from the maritime sector. Moreover, we use the interviews to determine the level of coverage of the ENISA scenarios and to gain insight in the role of threat information sharing.
 
\subsection{Method}
The interviews are based on a semi-structured questionnaire with questions in four different categories:
\begin{enumerate}
\item general context of the interviewee and their employer,
\item incident disclosure strategies and threat information sharing,
\item cyber incidents in the maritime sector in general,
\item the interviewees' perspective on the results presented in section~\ref{attacks}.
\end{enumerate} 

\subsection{General Context}
We interviewed representatives from five European companies from the maritime sector. Among them were logistic companies, shipyards and suppliers with a few hundred up to thousands of employees. Some of them are part of critical infrastructure. The interviewees were required to have insight in their company's IT security processes. Most have a full technical IT security position, some have to deal with organizational aspects of IT security and IT in general as well.
In order to guarantee the anonymity of the individual participants and companies, we will not release any more precise data.

All interviewees report that the number of cybersecurity full-time positions is small (one to ten employees). The number of employees who have insight into the cyber attacks the company is confronted with is always strictly limited and ranges from three up to 45 people. Most of the interview partners stated, that their companies see IT security as a relevant topic and have advanced attack detection methods that go beyond firewalls.
Two companies have dedicated teams for OT security, which closely work together with the IT security teams.

\subsection{Thread Information Sharing}
Only the critical infrastructure companies report incidents in order to comply with reporting obligations under law. They don't have additional internal rules for reporting. Two of the interviewees know unpublished cyber incidents from other companies.

The main reason for not publishing incidents are fear of image loss and economic damage. Two companies participate in Threat Information Sharing (TIS) platforms and gained useful information in the past. The other three agree that it could be useful to participate in TIS groups.

Reasons for not-participating are the inability to contribute own incidents or distrust against the other TIS participants. 
Additionally, the information gain from TIS groups may be bigger, when the group is small, non-official and communicates verbally and informal.
To increase the efficiency of TIS groups, one interviewee proposed that all incidents should be reported anonymously. This would help to prevent image losses. Another suggestion was to filter and categorize the threat information to allow a more targeted and customized stream of information.

\subsection{Cyber Attacks on the Maritime Sector}
The interviewees state that from an attacker's persepective the most interesting attacks are media-effective or mass attacks like email campaigns. The NotPetya attack on Maersk~\cite{2018-Safety_4_S-Maersk_Lin} was seen as the worst incident in the maritime sector. In line with this, the most serious type of attacks in the last years are ransomware attacks.

For defence, the companies take measures to protect endpoints but also sensitize their employees because they are a key entry point for malware. 
Only two of the companies keep statistics on IT security incidents.

\subsection{Comparison of Experience with our Results}
The interviews put the results of our research in perspective. The first point of criticism against our statistic is the differentiation between the threats malware, targeted attacks, ransomware and DDoS attacks because the transitions are fluent and not clear at first glance. This emerged from the fact, that the public sources are often vague and do not describe all the causes and consequences exactly. Therefore it is hard to decide, in which threat category they belong. Another point of criticism is, that we may have overseen the threats CEO-fraud and malware infections via USB flash drive. We don't have examples for real world incidents like this, but they occur in the wild. We covered CEO-fraud under the threat category social engineering in section \ref{phishing} and the USB-attack under different categories, mostly malware, because first and foremost this is an infection method and we do not distinguish according to this. Additionally we cover attacks like AIS-spoofing under the category geo-localisation signals spoofing and jamming in section \ref{spoofing}.
For a comprehensive analyses of attacks, we include proof of concepts in our work, as they are possible attacks and should therefore be taken into account, even though we have not yet observed any matching real world attacks. 
According to the statement of two interviewees, we may cover maximum ten percent of all cyber incidents in the maritime sector. Nevertheless, all participants agree, that the distribution may be correct taking into account, that the attack distribution has changed in recent years. It should also be emphasized that reporting an incident to an authority or a security company doesn't imply that they occur in public statistics. On the contrary, they are mostly confidential. However for informational and advertising purposes, these institutions sometimes publish anonymized statistics about cyber security incidents, that they handled. And so they found there way into our evaluation.
Lastly, most of our participants agree that it would help a lot if more incidents were published. Not least by increasing the awareness against cyber attacks, that is currently low.

\section{Conclusion}
\label{conclusion}
Our survey shows that more and more information security incidents are reported publicly, which also correspondents to a general trend in any other industry. But in absolute terms, considering the size of the maritime sector and estimates of unreported incidents, still very few incidents are reported. The experts we interviewed suspect that the number of successful attacks we found covers 10 percent at maximum. 

Nowadays major malware campaigns delivered through email can hit a maritime supplier in the same way as they affect any other industry. 
This is amplified by the fact that there are more opportunities for malware to spread because even systems that used to be OT-based, such as ships, now have enough common IT infrastructure on board to be infected by standard malware.
We can see this trend in the distribution of threats seen in the maritime sector over the last twenty years.
More untargeted and simple malware attacks took place then, whereas more ransomware and targeted attacks take place nowadays.

Most incidents we found belong to one of two kinds of attacks: First, there are attacks that target companies and vulnerable systems in general, and not specifically the maritime sector, for example to make money via ransomware. Second, there are targeted attacks in which specific maritime companies, such as shipyards, are targeted for presumably espionage purposes or other political reasons. In most of these cases it is suspected that state sponsored actors may be involved, but in most cases there are no concrete public available evidences. 

In general the attribution for cyber attacks is difficult because real evidence is rarely presented. There may be political reasons to present a specific actor as attacker or there may be a desire to disguise a configuration error. For this survey, we have taken the information about the presumed attackers from the respective sources. These information should be taken with a grain of salt.

There are few incidents where companies proactively publish information voluntary or under duress. Often, the information is not detailed. Rarely do companies themselves or insiders report in particular detail on the cause and consequences of attacks.
This is among other things because the companies concerned are afraid of bad publicity and competitive disadvantages.
Detailed reports are particularly valuable because everyone can learn from them: the reports show what can happen, what went poorly and what went well in responding to incidents.

All public information about incidents can be used by all companies to improve their IT security. There are concerns, that attackers could also use these reports to attack other companies in the described way, which can increase an already existing attack vector. We argue that such report are more important in combating an attack than in exacerbating it because a successful attacker could already potentially attack anyone that uses similar infrastructure if the attack vectors remain unfixed. 
We have highlighted particularly noteworthy reports in this paper and believe that it is very important that companies continue to publish both prevented and successful attacks in order to help other companies, independent of their industry, to avoid the same kind of attacks and to increase the overall awareness.

Our interviews suggest that the awareness of attacks has already increased in large companies today whereas small companies tend to consider themselves unattractive to attackers. Yet, for many companies, regardless of their size, there is a tough trade-off between IT security and cost. Policy makers in various countries support this decision with requirements laid down in legislation. Also, while some of our interviewees see the possibility of Thread Information Sharing (TIS) as potentially helpful for attack prevention, there are several reasons not to use this tool. These include distrust of other TIS participants and fear of bad publicity. However, there are also ideas to make TIS more attractive, for example by making the participants anonymous.

Overall, the threats for participants of the maritime sector are rising and therefore the individual parties should cooperate to hold out against these threats. The attackers learned already how to cooperate and the attack vectors are growing constantly.

\begin{acks}
The authors would like to thank all experts who were available for interviews.
This work is partly funded by the German Federal Ministry of Transport and Digital Infrastructure as part of the IHATEC program for the development of innovative port
technologies.
\end{acks}

\bibliographystyle{ACM-Reference-Format}
\bibliography{main}

\onecolumn
\appendix
\section{List of publicly reported incidents}
\label{tab:incidents}
\begin{center}
\begin{longtable}{ | p{2.5cm}  | c | p{5.7cm} | p{2.8cm} | c | }\hline
\textbf{Threat Category} & \textbf{Year} & \textbf{Target} & \textbf{Threat Actor} & \textbf{Source} \\\hline
\endfirsthead
\hline\textbf{Threat Category} & \textbf{Year} & \textbf{Target} & \textbf{Threat Actor} & \textbf{Source} \\\hline
\endhead
\hline
\endfoot
\hline
\endlastfoot

\multirow{17}{*}{\parbox{2.5cm}{Abuse and Theft\\ of Data}}
& 2012 & Australian Customs and Border protection & Crime Syndicate & \cite{2018-Yachting_P-Could_you}\\
& 2013 & General Dynamics &  & \cite{2013-The_Washin-Confidenti}\\
& 2014 & Unknown &  & \cite{2014-Cyberkeel-MARITIME_C}\\
& 2016 & US Navy Subcontractor &  & \cite{2016-Reuters-Personal_d}\\
& 2016 & DCNS &  & \cite{2016-Reuters-Frances_D}\\
& 2016 & Shipping company &  & \cite{2016-businessin-High-tech}\\ 
& 2016 & Ports & BRpsd & \cite{2016-ICS-CERT-Navis_WebA}\\
& 2016 & DSME & North Korea & \cite{2017-Reuters-North_Kore}\\
& 2017 & Australian Defence Subcontractor &  & \cite{2017-iTnews-Hacked_Aus}\\
& 2017 & Clarksons &  & \cite{2018-Clarksons-Update_on}\\
& 2017 & Svitzer &  & \cite{2018-ABC_News-Svitzer_em}\\
& 2018 & US Navy Subcontractor & China & \cite{2018-The_Telegr-Chinese_ha}\\
& 2018 & Austal &  & \cite{2018-The_Guardi-Defence_sh}\\
& 2019 & Holland America Line &  & \cite{2020-prnewswire-Holland_Am}\\
& 2019 & London Offshore Consultants & Maze Group & \cite{2020-Splash-London_Off}\\
& 2020 & DSME &  & \cite{2020-Twitter-Threat_act}\\
& 2021 & Carnival &  & \cite{2021-Bleeping_C-Carnival_C}\\
\hline

Data Manipulation & 2014 & Massachusetts Maritime Academy & Moroccan Electronic Islamic Union & \cite{2014-CyberKeel-Marine_Cyb}\\
\hline

\multirow{4}{*}{\parbox{2.5cm}{DDoS}}
& 2001 & Port of Houston & UK Teenager & \cite{2003-The_Regist-UK_teenage}\\
& 2001 & Port of Houston & US Citizen & \cite{2003-The_Regist-UK_teenage}\\
& 2013 & Port of Long Beach &  & \cite{2018-cytegic-Managing_t}\\
& 2017 & Port of Vancouver &  & \cite{2017-Columbian-Port_of_Va}\\
\hline
\multirow{11}{*}{\parbox{3cm}{Geo-localisation \\Spoofing}}
& 2001 & Unspecified & Proof of Concept & \cite{carroll2003vulnerability}\\
& 2009 & Unspecified & Proof of Concept  & \cite{2009-Journal-GPS_jammin}\\
& 2010 & South Korea & North Korea & \cite{GPSVuln13}\\
& 2012 & Unknown & North Korea & \cite{2013-Reuters-Government}\\
& 2013 & A single yacht & Proof of Concept & \cite{2013-Cockrell_s-UT_Austin}\\
& 2013 & Ship owners, Ship operators &  Proof of Concept & \cite{2014-Cyberkeel-MARITIME_C}\\
& 2013 & Generally Ships & Proof of Concept  & \cite{2014-Technical_-A_wake-up}\\
& 2016 & Generelly Ships &  & \cite{2016-Reuters-South_Kore}\\
& 2017 & Generally Ships &  & \cite{2017-The_Mariti-Mass_GPS_S}\\
& 2018 & Commercial Vessels &  & \cite{2019-US_Depar-MSCI_Advis}\\
& 2018 & Gernerally Ships & Proof of Concept  & \cite{2018-Riviera-Bridge_sys}\\
& 2018 & Generally Ships & Proof of Concept  & \cite{2018-Wilfried_H-Polizei-In}\\
& 2019 & Generally Ships &  & \cite{2019-MIT_Techno-Ghost_ship}\\
\hline

\multirow{2}{*}{\parbox{3cm}{Interception of \\emissions}}
 & 2014 & World Fuel Services &  & \cite{2014-Ship_and_B-WFS_In_Cou}\\
& & & &\\
\hline

\multirow{16}{*}{\parbox{2.5cm}{Malware}}
& 2013 & Mobile Offshore Drilling Unit &  & \cite{2015-Maritime_C-Coast_Guar}\\
& 2017 & JNPT & Sandworm & \cite{2017-Healthcare-Nuance_kno}\\
& 2017 & Maersk Line, APM Terminals, Damco & Sandworm & \cite{2018-Safety_4_S-Maersk_Lin}\\
& 2017 & Port of Rotterdam & Sandworm & \cite{2017-NL_Times-Rotterdam}\\
& 2018 & Cosco & Sandworm & \cite{2018-Safety_4_S-Cyber_atta}\\
& 2018 & Port of San Diego & Iran & \cite{2019-San_Diego_-What_happe}\\
& 2018 & Naval Industry &  & \cite{2019-Yoroi-The_Story}\\
& 2019 & Pitney Bowes &  & \cite{2019-Security_W-Pitney_Bow,2019-Pitney_Bow-System_Upd}\\
& 2019 & A single vessel &  & \cite{2019-James_Rund-Coast_Guar}\\
& 2019 & Ports & Proof of Concept  & \cite{2019-CyRiM_Repo-Shen_attac}\\
& 2019 & MTSA regulated facility &  & \cite{2019-US_Coast-Marine_Saf}\\
& 2020 & Toll Group &  & \cite{2020-CISOMAG-Mailto_Ran}\\
& 2020 & CGA CGM &  & \cite{2020-SWZ_Mariti-CMA_CGM_su}\\
& 2020 & MSC Cargo &  & \cite{2020-Security_W-Shipping_G}\\
& 2020 & Blue Water Shipping &  & \cite{2020-ShippingWa-Blue_Water}\\
& 2020 & Carnival &  & \cite{2020-Carnival-Substitute,2020-Bleeping_C-Worlds_la}\\

\multirow{4}{*}{\parbox{2.5cm}{Malware}}
& 2020 & Anglo-Eastern &  & \cite{2020-Splash-Anglo-East}\\
& 2020 & Hurtigruten &  & \cite{2020-Security_W-Norwegian}\\
& 2020 & Toll Group &  & \cite{2020-IT_News-Toll_Group2}\\
& 2021 & Rubin Design Bureau & Chinese APT & \cite{2021-The_Hacker-New_Chines}\\
\hline

\multirow{2}{*}{\parbox{2cm}{Man in the \\Middle}}
 & 2013 & Container carriers & Proof of Concept  & \cite{2014-Cyberkeel-MARITIME_C}\\
 & 2015 & Container carriers & Proof of Concept  & \cite{2015-Lloyds_Loa-Top_contai}\\
\hline

\multirow{15}{*}{\parbox{2.5cm}{Social Engineering, Phishing and \\Identity Theft}}
& 2013 & Major Shipping \& Trading Inc &  & \cite{2018-Kennedys-Whaling}\\
& 2014 & Container carriers & Proof of Concept & \cite{2014-CyberKeel-Marine_Cyb}\\
& 2015 & Shipping company &  & \cite{2016-American_a-MARITIME_C}\\
& 2015 & Nautilus Minerals, Marine Assets Corporation &  & \cite{2015-Insurance_-Cyber_hack}\\
& 2016 & Seafarer &  & \cite{2016-Paul_Berri-Honeytraps}\\
& 2016 & Shipping company &  & \cite{2016-ClearSky-Business_E,2016-American_a-MARITIME_C}\\
& 2017 & Shipping company &  & \cite{2018-Securework-Gold_Galle}\\
& 2018 & Bunker Fuel Company &  & \cite{2017-The_Sun_Da-Kedah_bunk}\\
& 2019 & Commercial Vessels &  & \cite{2019-US_Coast-Marine_Saf}\\
& 2020 & Tug Operating Organization &  & \cite{2020-Cybersecur-Cyber_Thre}\\
& 2020 & Maritime Sector &  & \cite{2020-US_Coast-Marine_Saf}\\
& 2021 & Ship broker; Bunker Supplier &  & \cite{2021-ITIC_Press-Counterfei}\\
& 2021 & MSC Cargo; Shipping company &  & \cite{2021-SAFETY4SEA-Recent_mar}\\
& 2021 & HAMMONIA Schiffsholding AG &  & \cite{2021-HAMMONIA_S-Finanziell}\\
& 2021 & Shipping company &  & \cite{2021-ITIC_Press-The_phone}\\
\hline

\multirow{16}{*}{\parbox{2.5cm}{Targeted Attack}}
& 2011 & Port of Antwerp & Crime Syndicate & \cite{2013-BBC-Police_war}\\
& 2011 & Shipbuilding, Maritime Operations & Icefog & \cite{2013-Kaspersky-The_ICEFOG}\\
& 2011 & Islamic Republic of Iran Shipping Lines &  & \cite{2012-Reuters-Irans_top}\\
& 2012 & Danish Maritime Authorities & China & \cite{2014-ShippingWa-Kina_hacke}\\
& 2012 & transportation companies &  & \cite{2014-Cylance-Operation}\\
& 2013 & Hyundai Merchant Marine & Kimsuky & \cite{2013-Kaspersky_-The_Kimsu}\\
& 2014 & Logistic companies &  & \cite{2014-TrapX_Secu-Anatomy_of}\\
& 2014 & Maritime Sector & Leviathan & \cite{2017-Proofpoint-Leviathan}\\
& 2015 & Maritime Sector & Leviathan & \cite{2015-Spiceworks-Stealthy_C, 2018-FireEye-Suspected}\\
& 2017 & A single vessel &  & \cite{2017-Safety_at_-Shipping_m}\\
& 2018 & Shipbuilding &  & \cite{2018-Yoroi-Cyber-Espi}\\
& 2018 & Government institutions & Carbanak Group & \cite{2018-Nextgov-Russia_Lau}\\
& 2018 & Shipbuilder & Leviathan & \cite{2018-FireEye-Suspected}\\
& 2018 & Government institutions & Gamaredon Group & \cite{2018-Nextgov-Russia_Lau}\\
& 2019 & Hamburg shipping company &  & \cite{2019-Paloalto_N-xHunt_Camp}\\
& 2020 & Ports &  & \cite{2020-Reuters-Iran_says}\\
& 2020 & Hamburger logistic company &  & \cite{2020-Spiegel_on-Polizei_un}\\
\hline

\multirow{8}{*}{\parbox{2.5cm}{Unknown Threat}}
& 2015 & TKMS, DCNS, Mitsubishi, Kawasaki HI &  & \cite{2015-Business_I-Foreign_bu}\\
& 2017 & BW Group &  & \cite{2017-SP_Global-Shipping}\\
& 2018 & Port of Barcelona &  & \cite{2018-Ara-El_port_de}\\
& 2019 & James Fisher and Sons &  & \cite{2019-The_Mariti-James_Fish}\\
& 2020 & Shahid Rajaei Port &  & \cite{2020-Washington-Officials}\\
& 2020 & International Maritime Organization &  & \cite{2020-Reuters-UN_shippin}\\
& 2021 & Bourbon &  & \cite{2021-Splash247-Bourbon_co}\\
& 2021 & Hyundai Merchant Marine &  & \cite{2021-Splash247-HMM_confir}\\
\hline

\multirow{6}{*}{\parbox{2.5cm}{Vulnerabilities}}
& 2013 & Lockheed Martin & Proof of Concept & \cite{2013-Reuters-Cyber_vuln}\\
& 2014 & US Navy & Proof of Concept  & \cite{2014-CyberKeel-Marine_Cyb}\\
& 2015 & Maritime Sector & Proof of Concept   & \cite{2015-Splash-Lack_of_pa}\\
& 2017 & A single vessel & Proof of Concept & \cite{2017-Pen_Test_P-OSINT_from}\\
& 2017 & VSAT systems & Researcher & \cite{2017-Mfame-Remote_Att}\\
& 2020 & A single vessel & Researcher & \cite{2020-Ewan_Robin-Hacked_-_a}\\
\hline
\end{longtable}
\end{center}

\end{document}